\documentclass[conference]{IEEEtran}
\IEEEoverridecommandlockouts
\usepackage{cite}
\usepackage{amsmath,amssymb,amsfonts}
\usepackage{algorithmic}
\usepackage{graphicx}
\usepackage{textcomp}
\usepackage{xcolor}
\usepackage[utf8]{inputenc}
\usepackage{array}
\usepackage{multirow}
\usepackage{tikz}
\usetikzlibrary{calc}
\usepackage[capitalise,nameinlink]{cleveref}
\usepackage{subcaption}
\usepackage{lipsum} 
\usepackage{eso-pic} 

\def\BibTeX{{\rm B\kern-.05em{\sc i\kern-.025em b}\kern-.08em
    T\kern-.1667em\lower.7ex\hbox{E}\kern-.125emX}}
\begin{document}

\title{Self-heating effect in nanoscale SOI Junctionless
FinFET with different geometries\\
\thanks{978-1-6654-4452-1/21/\$31.00 ©2021 IEEE.}
}

\author{\IEEEauthorblockN{A.E. Atamuratov, B.O. Jabbarova, M.M. Khalilloev}
\IEEEauthorblockA{\textit{Physics department} \\
\textit{Urgench State University}\\
Urgench, Uzbekistan \\
email: atabek.atamuratov@yahoo.com}
\and

\IEEEauthorblockN{A. Yusupov}
\IEEEauthorblockA{\textit{Department of Electronics and
Electrical Engineering} \\
\textit{Tashkent University of Information
technologies}\\
Tashkent, Uzbekistan\\
email: ayus@mail.ru}
\and
\IEEEauthorblockN{A.G.Loureriro}
\IEEEauthorblockA{\textit{Department.of Electronics and
Computer Sciences} \\
\textit{University of Santiago de Compostella}\\
Santiago de Compostella, Spain\\
email: antonio.garcia.loureiro@usc.es}
}

\maketitle

\begin{tikzpicture}[remember picture,overlay]
    \node[rotate=90, anchor=south] at ($(current page.west) + (1cm,0)$) {2021 13th Spanish Conference on Electron Devices (CDE) | 978-1-6654-4452-1/21/\$31.00 ©2021 IEEE | DOI: 10.1109/CDE52135.2021.9455728};
\end{tikzpicture}

\begin{abstract}
In this work we study the self-heating  effect (SHE) in nanoscale Silicon on Insulator Junctionless (SOI JL) FinFET transistor with fin cross section in rectangular, trapeze and triangle form. The lattice temperature dependence on the channel length as well as on buried oxide thickness is considered. It is shown that for considered transistor structure the lattice temperature in the middle of the channel is lower than at lateral sides, near source and drain. Also, we have found at the same conditions the lattice
temperature depends on shape of channel cross section too.
\end{abstract}


\section{Introduction}
The scaling of the metall-oxide-semiconductor
field effect transistors (MOSFET) is one of the main trends
of electronics based on CMOS technology. Decreasing of
the transistor sizes leads to decreasing of the consuming
energy and increasing the integration degree in integral
circuit. However, alongside with it the different
degradation effects such as short channel effects \cite{bib1},
decreasing of reliability \cite{bib2} and SHE \cite{bib3} can appear. For
decreasing the short channel effects SOI technology and
multigate FinFET, particularly Junctionless FinFET
structure were proposed \cite{bib1}. However, the low thermal
conductivity of buried oxide induces SHE in the channel
and decreasing the drain current.

Besides it, at manufacturing nanoscale transistors, in
technological processes, can be observed deviations of
geometrical sizes from stipulated one. For FinFET this
deviations can lead to change the stipulated transistor
channel shape \cite{bib4}. Therefore in this work SHE in SOI JL
FinFET with different geometries is analyzed.

\section{SIMULATION PROCEDURE}

We have studied a SOI JL FinFET with a equivalent
thickness of $HfO_2$ gate oxide $t_{ox}$=0.9 nm , and the width of

\begin{figure}[ht]
\centerline{\includegraphics[width=0.54\linewidth]{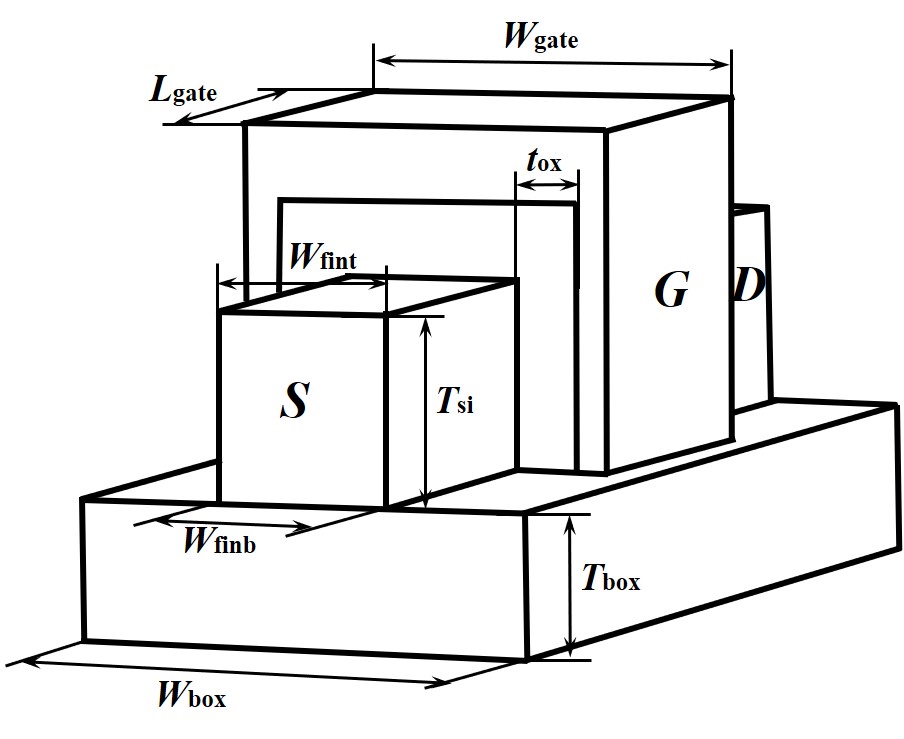}} \label{fig1a}
\centerline{\includegraphics[width=0.54\linewidth]{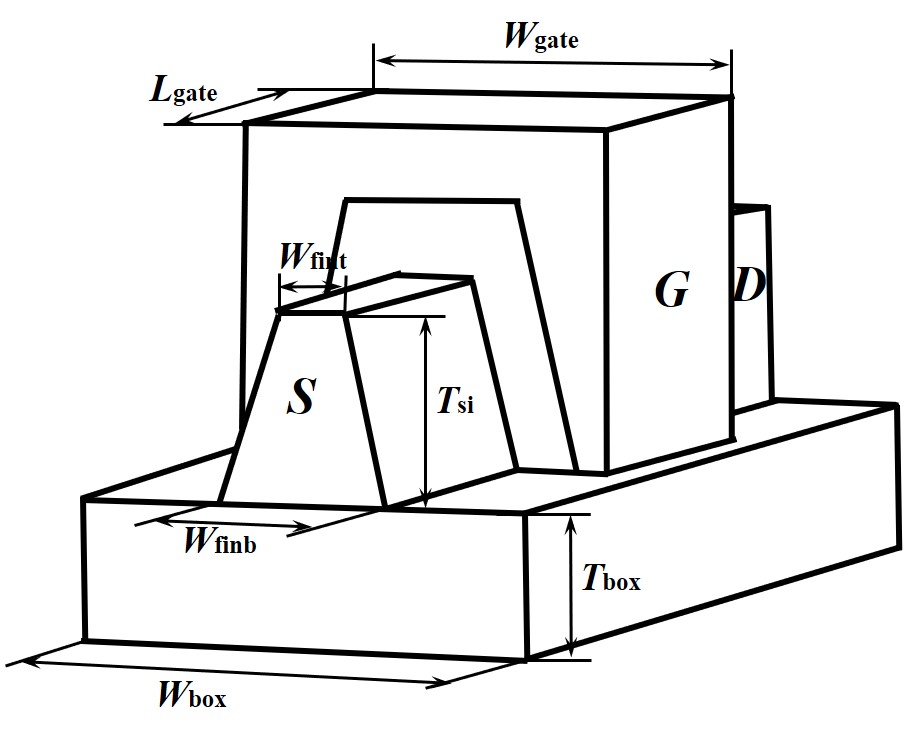}} \label{fig1b}
\centerline{\includegraphics[width=0.54\linewidth]{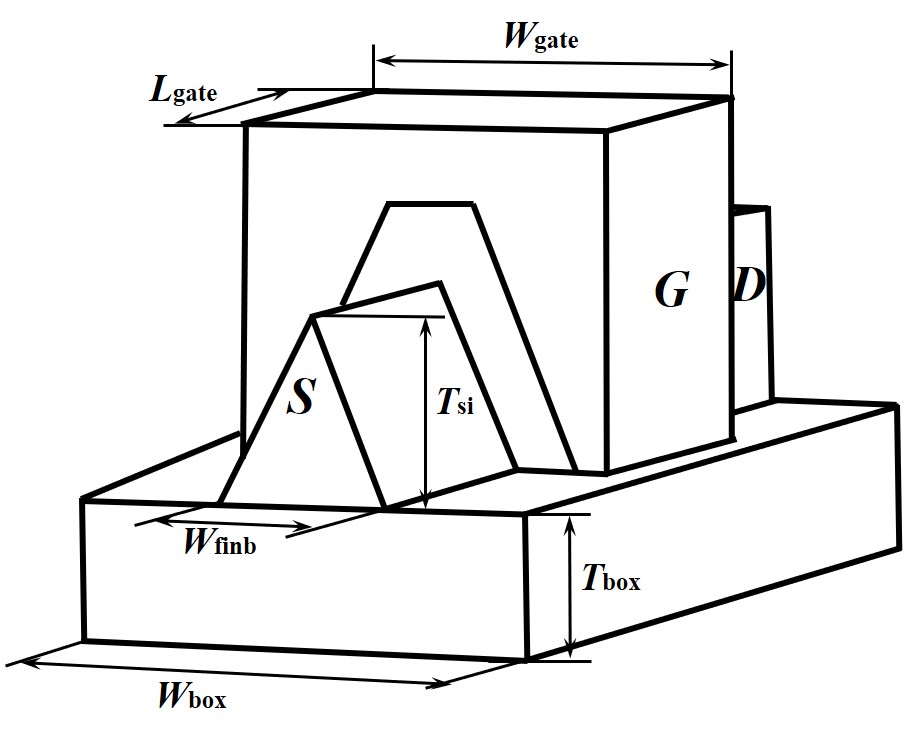}} \label{fig1c}
\caption{Id-Vg characteristics of inversion mode FinFET (1) and accumulation mode JL FinFET (2) with the same geometric parameters.}
\label{fig1}
\end{figure}

$SiO_2$ buried oxide (BOX) is $W_{box}$=69.4 nm. The thickness
of the BOX varies between $T_{box}$=10nm and 150 nm and
the length of TiN gate between $L_{gate}$=10nm and 40 nm.
Another main values are: the thickness of the n-Si channel
$T_{Si}$ =9 nm, width in the base $W_{finb}$ = 22 nm and width in the channel top depend on the channel shape, which for
considered trapeze cross section is $W_{fint}$=10 nm, and is zero
for triangle cross section (Figure 1).

\begin{figure}[ht]
\centerline{\includegraphics[width=0.9\linewidth]{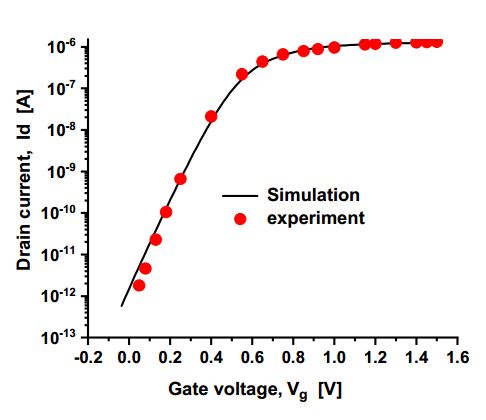}} 
\caption{Id-Vg dependence for simulated and experimental devices.
$V_{ds}$=0.9 V.}
\label{fig2}
\end{figure}

The device simulations were carried out using
Advanced Sentaurus TCAD. For device simulation, along
with default carrier transport model, mobility degradation model such as doping dependence to account
impurity scattering effect where considered. High field
saturation model were used to take into consideration
velocity saturation effect. Transverse field effect is
included to involve degradation at interfaces. To account
self-heating, a thermodynamic model for carrier transport,
SRH (temperature dependent) models were included.
Density gradient quantization model and mobility degradation due to high-k materials were also
taken into consideration. Figure 2 shows the very good
fitting of transfer characteristics of designed SOI JL
FinFET with experimental, presented in \cite{bib5}. Experimental
and simulated SOI JL FinFET with rectangular cross sections of
the channel have the following geometric parameters: gate length
$L_g$=13 nm, equivalent gate oxide thickness $t_{ox}$ =1.2 nm, height of
the channel $T_{si}$= 9 nm, width of the channel W=22 nm.

\section{SIMULATION RESULTS AND DISCUSSION}

In the Figure 3 the Id-Vg characteristics for SOI JL
FinFET with rectangular cross section, without and with
accounting self-heating effect is shown. Simulation results
shows, SHE in the nanoscale SOI JL FinFET induces
substantial decreasing of the drain current at high drain and
gate voltages. It is connected with influence of the
temperature induced by self-heating effect to mobility and
gate capacitance \cite{bib6}.

\begin{figure}[ht]
\centerline{\includegraphics[width=0.9\linewidth]{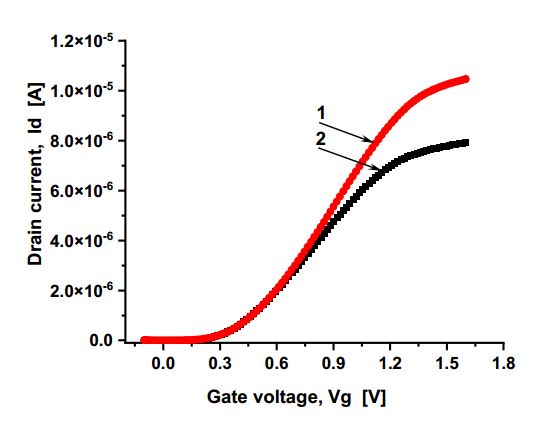}} 
\caption{Transfer characteristics of SOI JL FinFET with
rectangular cross section, without (1) and with (2) accounting the
self-heating effect. $V_{ds}$=0.75 V.}
\label{fig3}
\end{figure}

On the lattice temperature distribution along the device
length it is seen that at the middle of the length (in the
channel) the temperature is lower than in lateral parts, near
source and drain (Figure 4). It is due to different materials
which surround middle (gate oxide) and lateral parts (air).
Besides it, in the middle part along device length, in
bottom and top of the fin the temperature is lower than in
the middle of fin in vertical direction. It can be explained
by more fast heat transfer at bottom and top silicon-oxide
border. However, temperature dependence on position
which is seen in the vertical direction at lateral parts is
connected with more heat transfer through back oxide
(bottom part of fin) than through air (top part of fin). Small
decreasing of temperature at device edges is connected
with heat transfer through contacts.

\begin{figure}[ht]
\centerline{\includegraphics[width=0.9\linewidth]{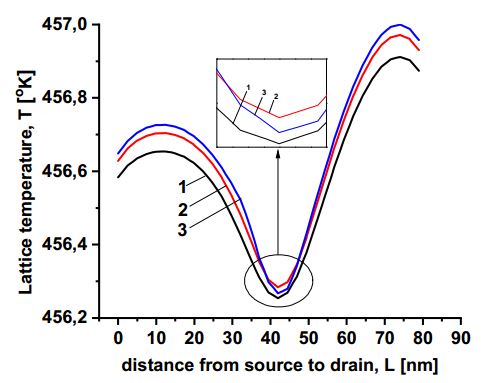}} 
\caption{Lattice temperature distribution along device with
rectangular cross section at bottom (1), middle (2) and top (3) of
the fin.}
\label{fig4}
\end{figure}

Figure 5 shows that lattice temperature is increased
linearly with increasing the gate length. It is due to contribution of field normal component connected with
gate voltage to electron kinetic energy. At increasing the
gate length the number of free electrons in the channel
covered by normal field is increased. At the same channel
height and the bottom width a volume (or area of cross
section), therefore the number of free electrons, of the
channel with rectanglular cross section is higher than for
channel with trapeze and triangle cross section. As
consequence the number of electrons with high kinetic
energy, therefore the channel temperature is high for
transistor with rectangle cross section.

\begin{figure}[ht]
\centerline{\includegraphics[width=0.9\linewidth]{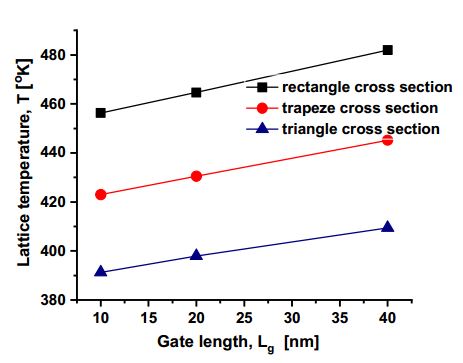}} 
\caption{Lattice temperature dependence on the gate length for
transistors with different cross sections. $T_{box}$=145 nm}
\label{fig5}
\end{figure}

\begin{figure}[ht]
\centerline{\includegraphics[width=0.9\linewidth]{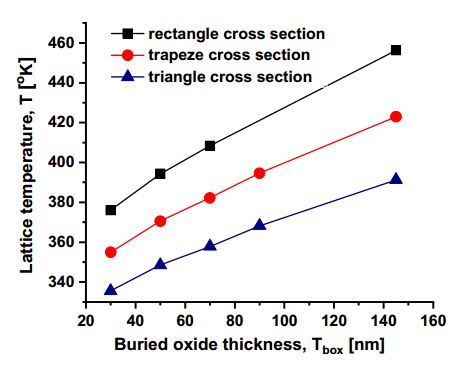}} 
\caption{Lattice temperature dependence on the buried oxide
thickness. $L_g$=10 nm}
\label{fig6}
\end{figure}

Lattice temperature dependence on the buried oxide
thickness is shown in the Figure 6. The lattice temperature
is linearly increased with an increase in buried oxide
thickness. It is agreed with model suggested in \cite{bib7}. In
accordance with this model the change of the lattice
temperature of the channel linearly depends on the thickness of buried oxide $T_{box}$ :

\begin{equation}
\Delta T = \frac{(P_t T_{box})}{(K_b A)} \label{eq1}
\end{equation}

where $K_b$ is heat conductivity of back oxide material, $P_t$ is
heat dissipation power, A is contact area to back oxide.
Increasing the lattice temperature with increasing the $T_{box}$
is caused by increasing the distance between channel and
thermal contact under buried oxide with increasing the
$T_{box}$. Therefore increasing the $T_{box}$ decrease the rate of
removing the heat from the channel.

\section{CONCLUSION}
From simulation study of SOI JL FinFET of
perspective of self heating it is observed that self heating
effect is different for transistors with rectangular, trapeze
and triangle cross sections at the same bottom width and
thickness of the channel. At the same conditions highest
lattice temperature in the channel is observed in transistor
with rectangular cross section. It is connected with larger
area for rectangular cross section. Lowest temperature
along channel is seen in the middle for all channel shapes.
Lattice temperature, also, depend on gate length $L_g$ and
buried oxide thickness $T_{box}$ and it is increased linearly with
increasing $L_g$ and $T_{box}$. Increasing the lattice temperature
with increasing the gate length is result of increasing the
area affected by normal gate field, which induce additional
kinetic energy for electrons in the channel. Linear
increasing the lattice temperature with increasing $T_{box}$ is

consequence of increasing the distance between channel
and contact under BOX.

\section{ACKNOWLEDGMENT}

This work has been supported by Ministry of
innovative development of the Republic of Uzbekistan
under contract Uzb-Ind-2021-80.

\bibliographystyle{unsrt}

\bibliography{ref}

\end{document}